\documentclass[preprint,aps,nofootinbib]{revtex4-1}

\usepackage{dcolumn}
\usepackage{bm}
\usepackage{epsfig}
\usepackage{graphicx}
\usepackage{float}
\usepackage{paralist}

\usepackage{amsmath}
\usepackage{amsfonts}
\usepackage{amssymb}

\usepackage{color}

\usepackage[utf8]{inputenc}
\usepackage{slashed}
\usepackage{indentfirst}
\usepackage{caption}
\usepackage{subcaption}
\usepackage{hyperref}

\begin{document}

\title{
Stop Search in the Compressed Region via Semileptonic Decays
}
\author{Hsin-Chia Cheng, Christina Gao, Lingfeng Li, Nicol\'as A. Neill}
\affiliation{Department of Physics, University of California, Davis, California 95616, USA}

\begin{abstract}
In supersymmetric extensions of the Standard Model, the superpartners of the top quark (stops) play the crucial role in addressing the naturalness problem. For direct pair-production of stops with each stop decaying into a top quark plus the lightest neutralino, the standard stop searches have difficulty finding the stop for a compressed spectrum where the mass difference between the stop and the lightest neutralino is close to the top quark mass, because the events look too similar to the large $t\bar{t}$ background. With an additional hard ISR jet, the two neutralinos from the stop decays are boosted in the opposite direction and they can give rise to some missing transverse energy. This may be used to distinguish the stop decays from the backgrounds. In this paper we study the semileptonic decay of such signal events for the compressed mass spectrum. Although the neutrino from the $W$ decay also produces some missing transverse energy, its momentum can be reconstructed from the kinematic assumptions and mass-shell conditions. It can then be subtracted from the total missing transverse momentum to obtain the neutralino contribution. Because it suffers from less backgrounds, we show that the semileptonic decay channel has a better discovery reach than the fully hadronic decay channel along the compressed line $m_{\tilde{t}} - m_{\tilde{\chi}}\approx m_t$. With 300 $\text{fb}^{-1}$, the 13 TeV LHC can discover the stop up to 500 GeV, covering the most natural parameter space region. 
\end{abstract}
\maketitle

\section{Introduction}
\label{sec:introduction}
The discovery of the Higgs boson~\cite{Aad:2012tfa,Chatrchyan:2012xdj} completes the Standard Model (SM), but also makes the hierarchy problem more eminent. The SM interactions of the Higgs field induce quadratically divergent contributions to its mass-squared, and the largest contribution comes from the top quark loop. In order to keep the electroweak symmetry breaking scale natural, new physics is expected to be present near the weak scale to cut off the divergent contributions. In supersymmetry (SUSY), the top quark loop is cancelled by the loops of its superpartners, the stops. It is hence natural that the stops belong to the most sought-after particles in new physics searches at the LHC. 

After Run 1 and the initial 13 TeV run of the LHC, ATLAS and CMS experiments have put constraints on the stop mass up to $\sim 750$~GeV, assuming the stop decays to a top quark and  the lightest neutralino $\tilde{\chi}_1^0$, which is assumed to be the lightest supersymmetric particle (LSP) and stable, with $m_{\tilde{\chi}_1^0} \lesssim 200$ GeV~\cite{Aad:2015pfx,Khachatryan:2016oia,Khachatryan:2016pup,CMS-PAS-SUS-16-002}.  Similar but slightly weaker bounds were also obtained if some of the stops decay through a chargino or a heavier neutralino to the LSP. Run 2 is expected to extend the reach beyond 1 TeV, at which point SUSY as a solution to the hierarchy problem may be strongly questioned. However, the current searches leave some gaps in the lower stop mass region. In particular, if $m_{\tilde{t}} \approx m_t + m_{\tilde{\chi}}$, the top quark and the neutralino from the stop decay are almost static in the stop rest frame. Consequently, in the lab frame the top and the neutralino will be collinear with $p_{\tilde{\chi}}/p_{\tilde{t}} \approx m_{\tilde{\chi}}/m_{\tilde{t}}$. In such cases, the stop pair production events will look almost identical to the top quark pair production, as the two neutralinos tend to travel back to back, resulting in a cancellation of their momenta and leaving no trace of $\tilde{\chi}$s. This is the reason why no experimental limit has been set upon this compressed region so far.

One possible way proposed in Refs.~\cite{Hagiwara,Liantao,Macaluso}  to explore the compressed region is to consider events of stop pair production with a hard initial state radiation jet ($J_{\text{ISR}}$). 
From momentum conservation, 
\begin{equation}\label{eq1}
p_{T}(J_{\text{ISR}}) \approx -(p_{T}(\tilde{t}_1) + p_{T}(\tilde{t}_2))\ , 
\end{equation}
therefore both neutralinos tend to be emitted in opposite direction to the ISR jet, resulting in a significant amount of missing transverse momentum ($\slashed{p}_T$). For the fully hadronic decay events, the $\slashed{p}_T$ mainly comes from the neutralinos. Using Eq.~(\ref{eq1}), we see that the ratio between $\slashed{p}_T$ and $p_{T}(J_{\text{ISR}})$ (defined as $R_M$ in Ref.~\cite{Liantao}) is roughly equal to the ratio between the neutralino and the stop masses, 
\begin{equation}\label{RM}
R_M\equiv\frac{\slashed{p}_T}{p_{T}(J_{\text{ISR}})}\approx \frac{m_{\tilde{\chi}}}{m_{\tilde{t}}}\,  ,
\end{equation} 
which is strictly between zero and one.
It can be a useful kinematic variable to distinguish the stop events where $R_M$ should be close to ${m_{\tilde{\chi}}}/{m_{\tilde{t}}}$ from the SM top background events where $R_M$ is expected to be close to zero~\cite{Hagiwara,Liantao,Macaluso}.

As for the semileptonic and dileptonic decays of the stops, $R_M$ becomes less informative if the neutrinos' contribution to $\slashed{p}_T$ cannot be separated from that of neutralinos. However, for semileptonic events, if one exploits the kinematics unique to the compressed region, it is possible to reconstruct the top quark that decays leptonically, hence retrieving a relation similar to Eq.~(\ref{RM}). Another benefit of requiring a lepton in the final states is that it vetoes  QCD backgrounds, which suffer from large uncertainties under high jet multiplicities. 

In this paper we demonstrate the reconstructions of semileptonic decays of the stop pair production in the compressed region, and show that it is very useful for stop searches. In Sec.~\ref{sec:kinematics}, we analyze the kinematics of the semileptonic events and discuss the reconstruction of missing transverse momenta from the neutrino and the neutralinos. In Sec.~\ref{sec:case} we describe in detail a search done for the benchmark mass point $m_{\tilde{t}}=400$ GeV, $m_\chi=226.5$ GeV as an example of our method. We also compare the significances obtained from our method and the search based on the hadronic final state at the same benchmark point. Sec.~\ref{sec:results} gives the results of our method for the stop search at LHC 13 TeV for an integrated luminosity of 300 $\textrm{fb}^{-1}$, with the stop mass ranging from 250 GeV to 600 GeV in the compressed region. The conclusions are drawn in Sec.~\ref{sec:conclusions}.

\section{Kinematics of semileptonic events}
\label{sec:kinematics}
In semileptonic decays of stop pairs, the main challenge is to separate the $\slashed{p}_T$ component due to neutralinos from that of the neutrino. When a hard ISR jet is present, from momentum conservation, the stop pair are boosted in the opposite direction to the hard jet. If $m_{\tilde{t}} \sim m_t + m_{\tilde{\chi}}$, as explained in the Introduction, the two neutralinos travel approximately along the direction of their mother particles. In the transverse plane to the beam direction, we expect the sum of the two neutralino transverse momenta to be in the opposite direction to the $\slashed{p}_T$ of the ISR jet, hence contributing only to $\slashed{p}^{\parallel}_T$, the component of missing momentum antiparallel to $J_{\text{ISR}}$ . For semileptonic events, where one stop decays hadronically and the other leptonically, the component of $\slashed{p}_T$ perpendicular to $J_{\text{ISR}}$ can roughly be attributed to the presence of the neutrino. The size of this component is given by
\begin{equation}\label{eq2}
\slashed{p}_T^{\perp}=\slashed{p}_T-\frac{(\slashed{p}_T\cdot p_{T}(J_{\text{ISR}}))p_{T}(J_{\text{ISR}})}{|p_{T}(J_{\text{ISR}})|^2}\; \left(=p_{T\nu}^{\perp}\right)\, ,
\end{equation}
which we will assume to be the neutrino momentum component perpendicular to the ISR jet in the transverse plane for the following analyses. Once the $J_{\text{ISR}}$ is identified, $p_{T\nu}^{\perp}$ is uniquely determined from the experiment.

We first consider the case $m_{\tilde{t}} = m_t + m_{\tilde{\chi}}$. For the leptonically decaying top quark, there are three mass-shell equations in addition to Eq.~(\ref{eq2}):
\begin{equation}\label{eq3}
p_\nu^2 = 0\, ,
\end{equation}
\begin{equation}\label{eq4}
(p_\ell+p_\nu)^2 = m_{W}^2 \, ,
\end{equation}
\begin{equation}\label{eq5}
(p_\ell+p_\nu+p_b)^2 = m_t^2\, .
\end{equation}
Given the measured momenta of the lepton and the $b$-jet, the three equations together with the  $p_{T\nu}^{\perp}$ allow us to solve for $p_\nu$.
Taking the differences of the 3 mass-shell equations followed by plugging in the $p_{T\nu}^{\perp}$ from Eq.~(\ref{eq2}), we can reduce them to one quadratic equation for $E_\nu$, the kinetic energy of the reconstructed neutrino.

The quadratic equation, if solvable, provides in general two different real solutions for $E_\nu$. We will discuss how we select the solution later. After $E_\nu$ is determined, we substitute it back into the original mass-shell equations, then the full momentum of the reconstructed neutrino can be retrieved. Finally, with the knowledge of $p_{T\nu}^{\parallel}$, the component of the neutrino momentum antiparallel to $p_{T}(J_{\text{ISR}})$, we can subtract the neutrino contribution from $\slashed{p}_{T}^{\parallel}$ and get a relation similar to Eq.~(\ref{RM}):
\begin{equation}
\bar{R}_M\equiv\frac{\slashed{p}_{T(\chi)}}{p_{T(J_{\text{ISR}})}}\approx\frac{\slashed{p}^{\parallel}_{T}-p^{\parallel}_{T\nu}}{p_{T(J_{\text{ISR}})}}\approx \frac{m_{\tilde{\chi}}}{m_{\tilde{t}}},
\end{equation} 
where we define the variable $\bar{R}_M$ as the modified $R_M$ adapted to the semileptonic decays.
With a set of proper kinematic cuts, a clear peak in the $\bar{R}_M$ distribution for the stop pair production can be identified, as we will show later.

\begin{figure}[th]
\captionsetup{justification=raggedright,
singlelinecheck=false
}
\centering
\begin{subfigure}[b]{0.4\textwidth}
\includegraphics[width=\textwidth]{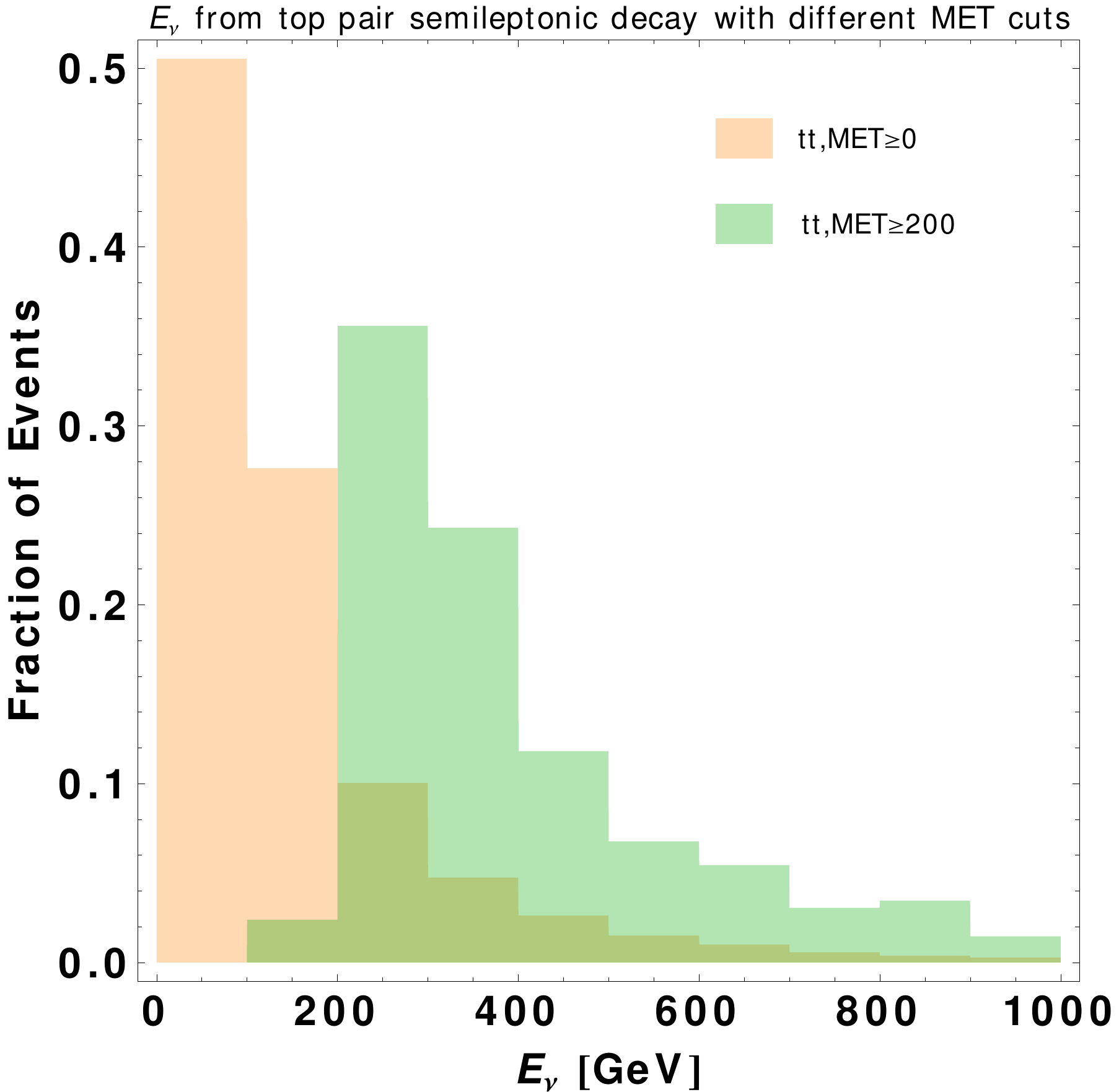}
\end{subfigure}
\begin{subfigure}[b]{0.4\textwidth}
\includegraphics[width=\textwidth]{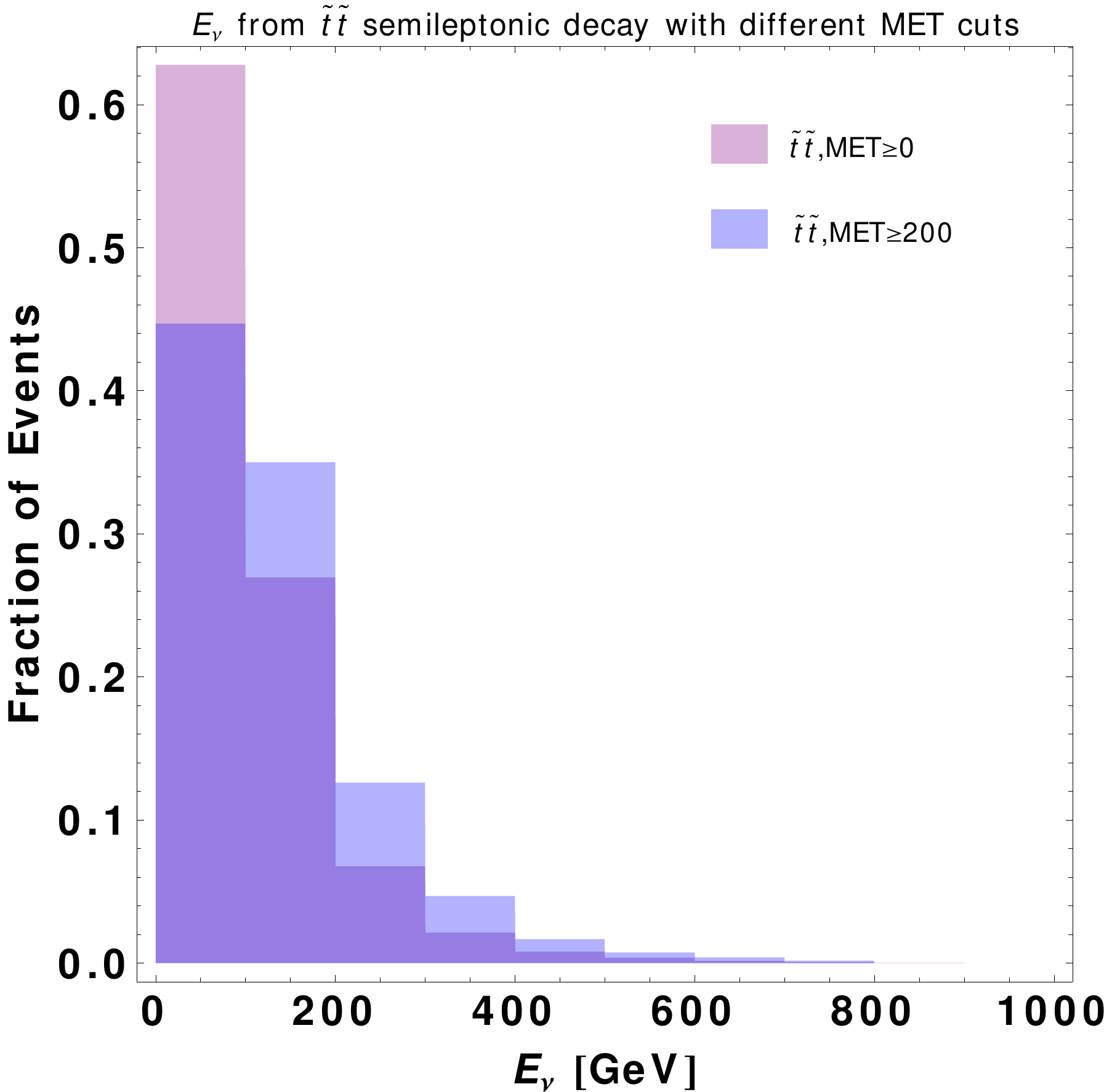}
\end{subfigure}

\begin{subfigure}[b]{0.4\textwidth}
\includegraphics[width=\textwidth]{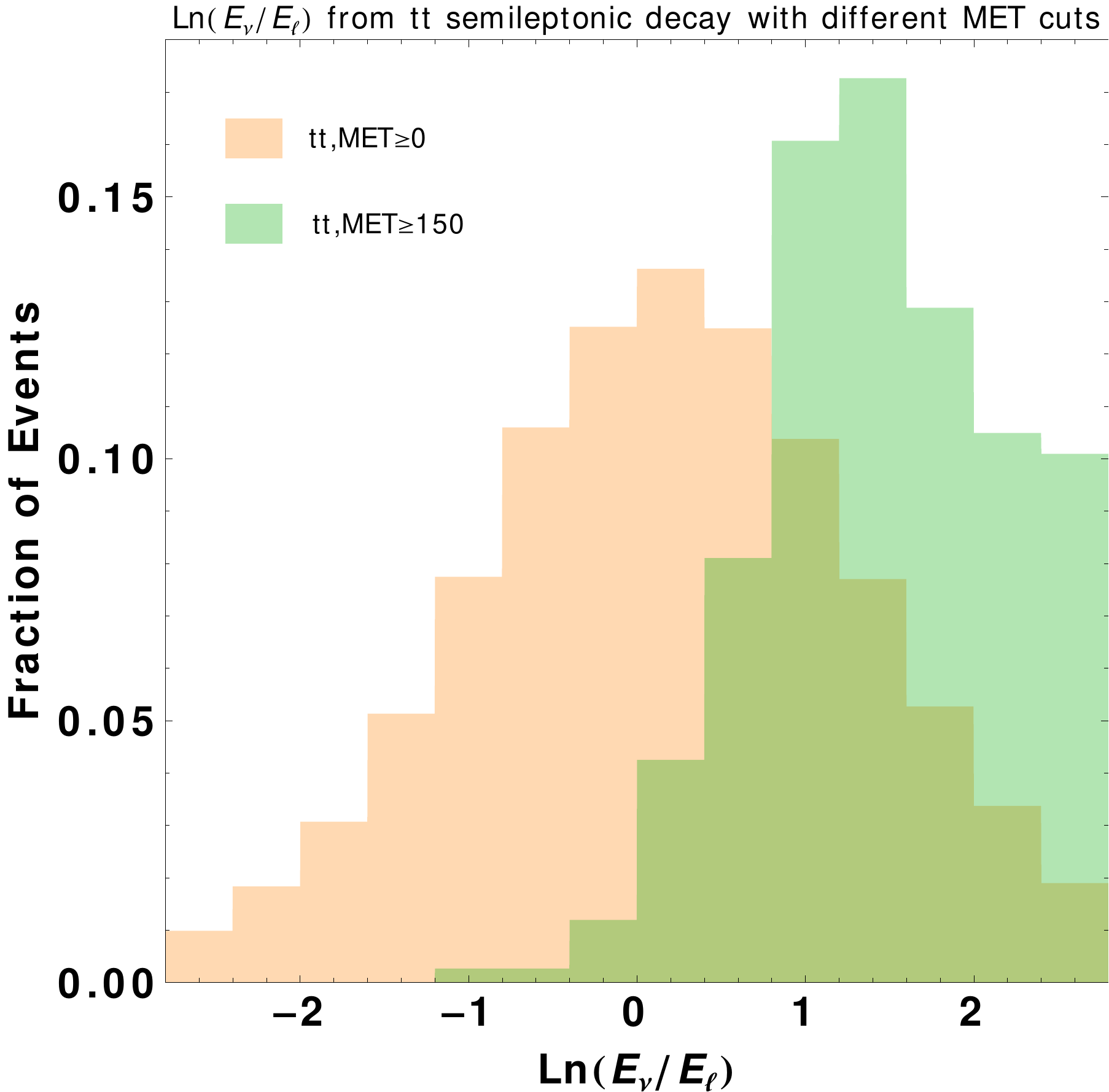}
\end{subfigure}
\begin{subfigure}[b]{0.4\textwidth}
\includegraphics[width=\textwidth]{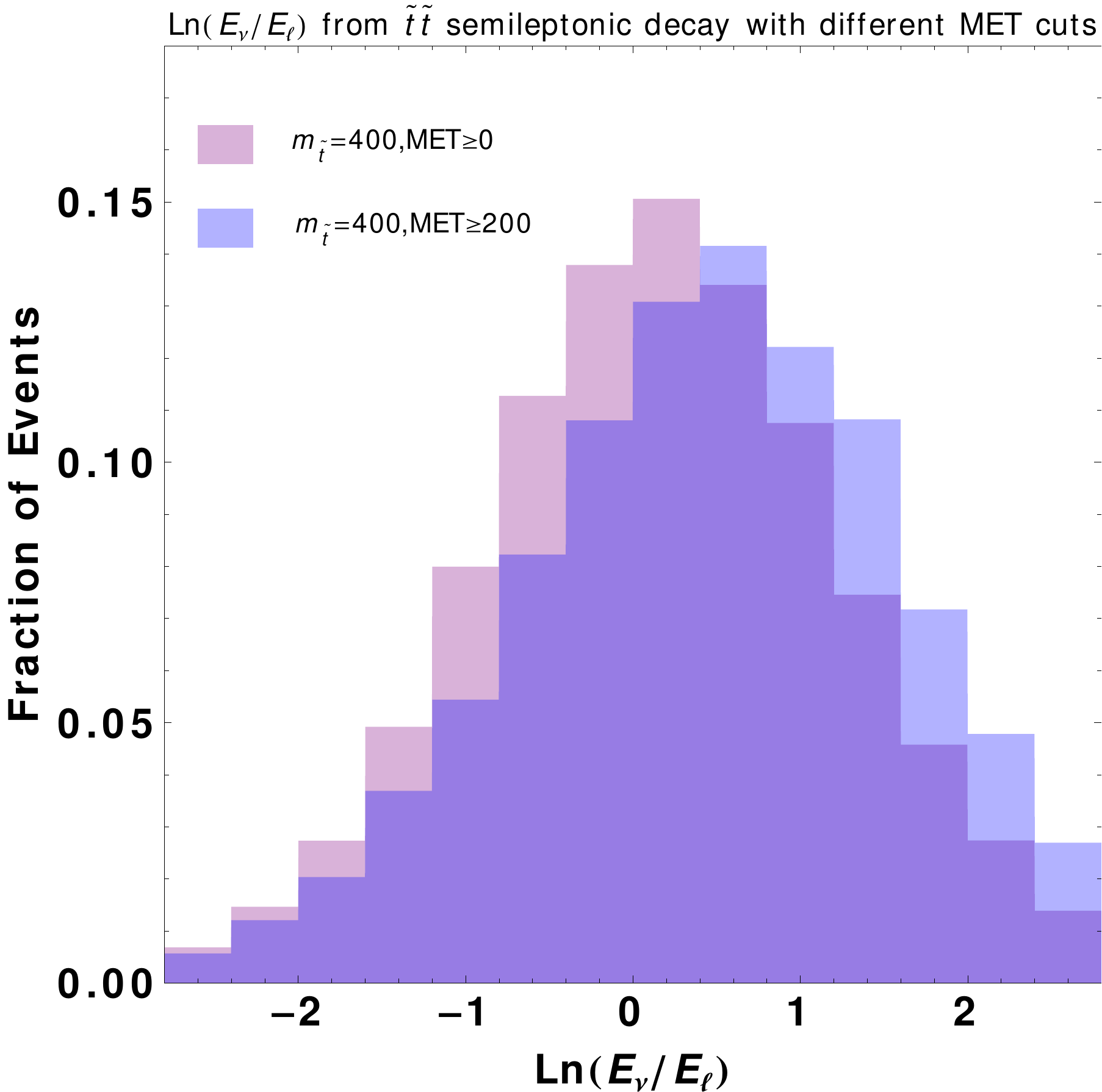}
\end{subfigure}
\caption{The comparisons of the parton level neutrino energy before and after a 200 GeV MET cut for both the stop and top pair production. All distributions are normalized to one. The minor asymmetry that appears in the $\log(E_\nu/E_\ell)$ distribution without the MET cut is because of the spin correlation between the neutrino and the $W$ boson.}
\label{met}
\end{figure}

As we discussed above, the quadratic equation in general can give two possible solutions for $E_\nu$. To choose between them, we investigate the kinematics of semileptonic decays for $\tilde{t}\bar{\tilde{t}}+ J_{\text{ISR}}$ and its main background $t\bar{t}+ J_{\text{ISR}}$. As an illustration, we generate these events at the parton level for a benchmark of $m_{\tilde{t}}=400$ GeV and $m_{\chi}=226.5$ GeV. The charged lepton and the neutrino from the $W$ decay on average have the same energy if the $W$ boson is longitudinally polarized, because they tend to be emitted in directions perpendicular to the $W$ momentum. On the other hand, for transverse $W$ decays the neutrino tends to be more energetic than the charged lepton. Because $W$ bosons coming from the top decays are dominantly longitudinally polarized, the energy distributions of the neutrino and the charged lepton are similar. We can see in Fig.~\ref{met} that the distribution of $\log(E_\nu/E_\ell)$ is quite symmetric around zero before any cut, with a slight bias towards the positive values due to the transverse $W$ contribution. The distributions are modified significantly after the missing transverse energy (MET) cut. In the upper panels of Fig.~\ref{met}, we show the MET distributions for the semileptonic decays of the $\tilde{t}\bar{\tilde{t}} + J_{\text{ISR}}$ signal and $t\bar{t}+J_{\text{ISR}}$ background. One can see that the stop events in general have a larger MET due to the presence of the neutralinos. 
Therefore, a MET cut can help to suppress the top background significantly. 
After a MET cut of 200 GeV, we see that the distribution of $\log(E_\nu/E_\ell)$ for the top background events is strongly shifted towards the positive values because most events with a small neutrino energy are discarded. For the stop signal, the distribution is also moved towards the positive values but the effect is less significant. 
Based on this observation, we will simply choose the solution with the greater neutrino energy of the two real solutions, with some upper limits which will be specified when we perform our case study.

For $W$ decaying to $\tau$ with a subsequent decay to an electron or a muon, two additional neutrinos are produced in the decay chain. The additional MET from the extra neutrinos makes the reconstruction of the correct neutrino momentum impossible, thus introducing an irreducible error into the distribution of $\bar{R}_M$. Fortunately, $\tau$ decays leptonically 35\% only, which makes this contamination rather small. On the other hand, if $\tau$ decays hadronically, with a charged lepton ($e$ or $\mu$) from the other $W$ decay, the $\nu_\tau$ associated with $\tau$ also gives additional MET. Such events may be partially removed by the $\tau$-tagging on the final states.

Besides issues discussed above, the limited jet and lepton energy resolutions at the detectors, the pile up effect and extra soft radiations can further smear the $\bar{R}_M$ distribution. Details about these effects are beyond the scope of this paper.

\section{A Case Study}
\label{sec:case}
As an illustration of our method, we describe in details the search done for a point $m_{\tilde{t}}=400$ GeV, $m_\chi=226.5$ GeV in the parameter space of the compressed region. 
The dominant SM backgrounds for the semileptonic decay of $\tilde{t}\bar{\tilde{t}}+ J_{\text{ISR}}$ are the semileptonic and dileptonic decays of $t\bar{t}+ J_{\text{ISR}}$. The reason that the dileptonic decays are important is mainly due to the imperfect lepton isolation. Since the top and its decay products are highly boosted antiparallel to the hard ISR jet, the lepton tends to have a small $\Delta R$ separation from the $b$ jet, therefore has a non-negligible probability of failing the lepton isolation criteria.
Both of these backgrounds have a similar topology to the signal, consequently they have a good chance of solving Eqs.~(\ref{eq2}) to (\ref{eq5}) and yielding a sensible $\bar{R}_M$ value lying between 0 and 1. 

Other SM backgrounds include the single or pair production of vector bosons ($V$) with jets and $t\bar{t}V$. Even though $V+\text{jets}$ and $VV+\text{jets}$ have relatively large cross sections, they seldom produce sensible solutions for the equations imposed by the signal kinematics. The small fractions that give real solutions rarely pass our selection cuts either. As a result, they give much less yields compared to $t\bar{t}+ J_{\text{ISR}}$. $t\bar{t}V$, on the contrary, has the kinematic features akin to the signal, but suffers from a tiny cross section. As a result, contributions of other SM backgrounds are negligible compared to the main backgrounds.
Because one isolated lepton is required in the final state, the hadronic decays of the top pair production and the pure QCD backgrounds are also negligible. 

Besides the SM backgrounds, the dileptonic decay of $\tilde{t}\bar{\tilde{t}}+ J_{\text{ISR}}$ can be an irreducible background to the signal. However, this process has a much smaller cross section compared to the SM processes and it is effectively negligible.

\subsection{Signal and background generations}
We use MadGraph 5~\cite{Madgraph1} and Pythia 8~\cite{Pythia1} to generate events for both the background and the signal events. MLM matching scheme is turned on to prevent double-counting between the matrix element calculation and the parton shower \cite{MLMmatching1}. The detector simulation is performed by Delphes 3~\cite{Delphes1} with the anti-$k_t$ jet algorithm~\cite{Antikt1}. 
We normalize the background cross sections to the LHC 13 TeV top production~\cite{TopXsession1,TopXsession2,TopXsession3,TopXsession4,TopXsession5}. A $K$-factor of 1.29 is applied to both semileptonic and dileptonic decays of the $t\bar{t}$ backgrounds. For the signals, the production cross section is normalized to LHC 13 TeV NLO+NLL results~\cite{StopXsession1}.

\subsection{Event selection}
The selection for the events of interest starts with at least 4 jets with one or more $b$-tags and exactly one isolated lepton. The $b$-tagging efficiency is set to be 80\% with a misidentification rate of 0.015~\cite{Btag1}. Events with $\tau$-tagging are vetoed. 
The non-$b$-tagged jet with the hardest $p_T$ is our ISR jet candidate. In particular, it must satisfy $p_T \geq 475$ GeV. The second and third hardest jets must satisfy $p_T \geq 60$ GeV. 
In order to ensure that the ISR jet is approximately in the opposite direction of the neutralino momentum sum, we require that $|\phi_{J_{\text{ISR}}}-\phi_{\text{MET}}| \geq 2$.
As shown in Fig.~\ref{met}, a MET cut effectively eliminates most of the SM backgrounds whose missing momentum mainly comes from the neutrinos, hence an MET cut $ >200$ GeV is imposed. 

For each $b$-tagged jet in the event that passes all the above preliminary selections, we check if Eqs.~(\ref{eq2}) to (\ref{eq5}) are solvable. Approximately 42$\%$ of the signal events give real solutions. For the semileptonic and dileptonic background, the fractions of events yielding real solutions are approximately 43$\%$ and 36$\%$, respectively. We then pick the higher neutrino energy solution for the reasons aforementioned in Sec.~\ref{sec:kinematics}. However, this choice is accompanied by the danger of accepting an unphysically large $E_\nu$. To avoid these unphysical solutions, upper limits on the reconstructed neutrino transverse momentum and the ratio are imposed. If a solution has $p_{T\nu}>$ 180 GeV or $p_{T\nu}/p_{T\ell}>6$, this combination of $b$-jet and lepton is discarded. For events with two $b$-tagged jets that allow for two $\bar{R}_M$ values, we select the smaller $\bar{R}_M$.

Another useful kinematic variable is $\Delta \phi_{\ell,\text{MET}}$, the azimuthal angle difference between the lepton and the missing transverse momentum. Since the main source of the missing momentum for the backgrounds is from the neutrino, it tends to be more collinear with the lepton for the background events compared with the signal events. %
\begin{figure}[t]
\captionsetup{justification=raggedright,
singlelinecheck=false
}
\begin{center}
\includegraphics[scale=0.4]{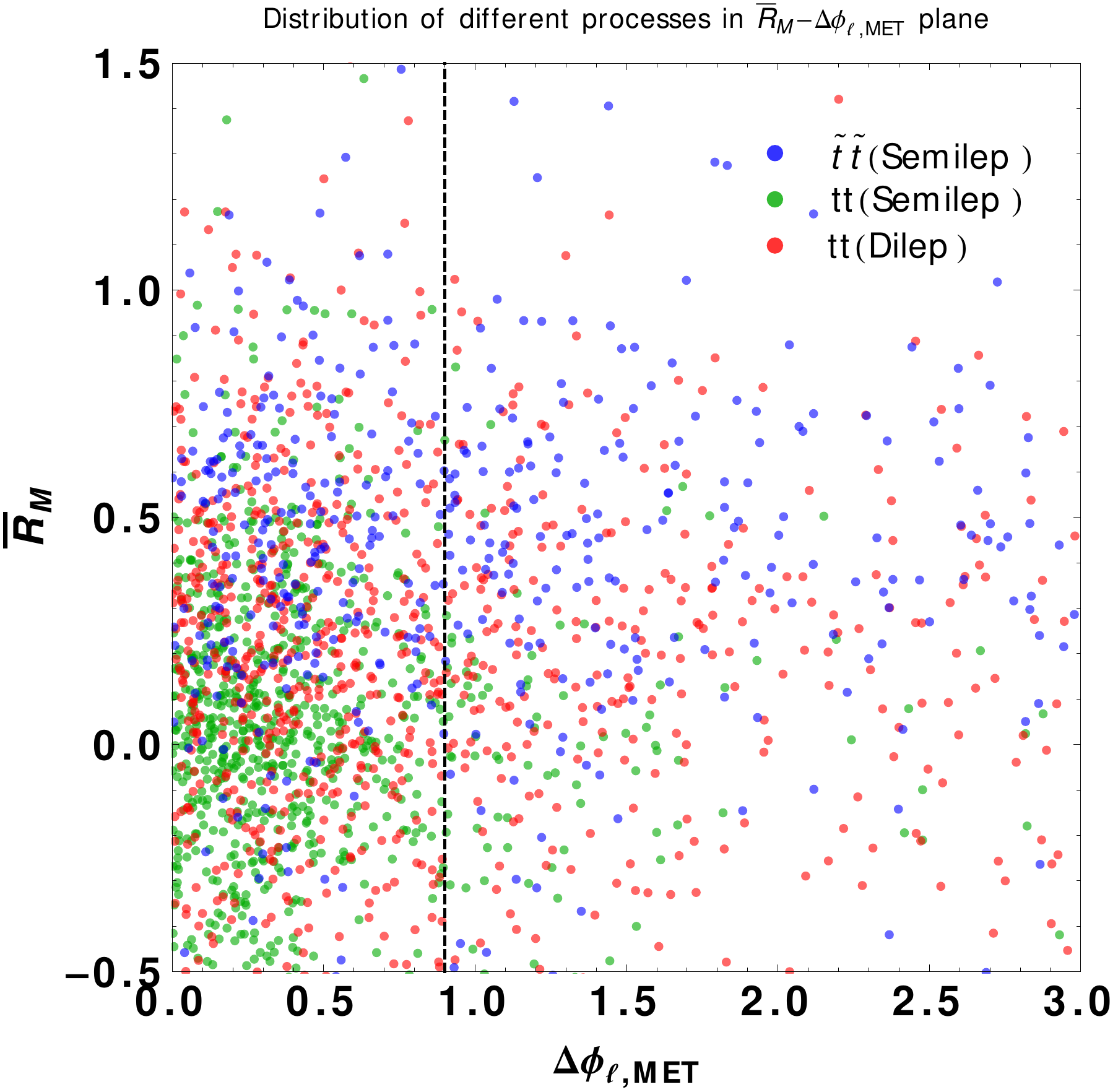}
\includegraphics[scale=0.4]{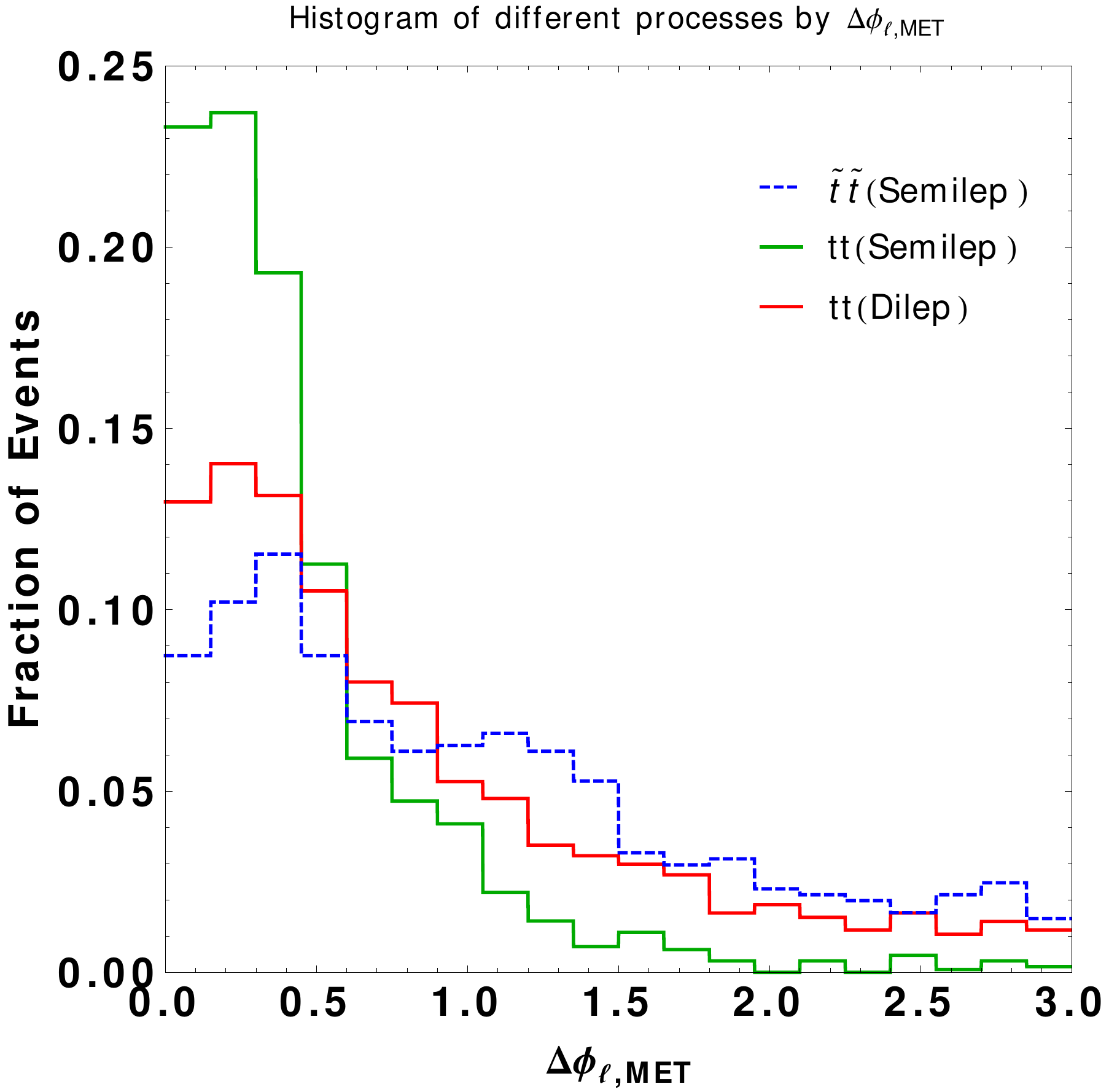}
\caption{$\Delta \phi_{\ell,\text{MET}}$ distributions for the signal and backgrounds. The semileptonic (dileptonic) decayed $t\bar{t}$ background is shown in green (red) points/curve. The signal is studied for the benchmark $m_{\tilde{t}}=400$ GeV and $m_{\chi}=226.5$ GeV, and is represented by blue points/curve. \textbf{Left:} Scattered plots of $\bar{R}_M$  vs. $\Delta \phi_{\ell,\text{MET}}$. The vertical line is $\Delta \phi_{\ell,\text{MET}}=0.9$. \textbf{Right:} Normalized distributions of $\Delta \phi_{\ell,\text{MET}}$.}
\label{deltaphi}
\end{center}
\end{figure}
Fig.~\ref{deltaphi} shows the $\Delta \phi_{\ell,\text{MET}}$ distributions for the signal and the backgrounds for the benchmark. After a cut on $\Delta \phi_{\ell,\text{MET}} >0.9$, most of the semileptonic background can be suppressed. However, this cut is less effective on the dileptonic background, because its MET is the sum of two neutrinos' momenta, which results in a wider $\Delta \phi_{\ell,\text{MET}}$ distribution.

\subsection{Results of the case study}
\begin{figure}
\captionsetup{justification=raggedright,
singlelinecheck=false
}
\centering
\begin{subfigure}[b]{0.4\textwidth}
\includegraphics[width=\textwidth]{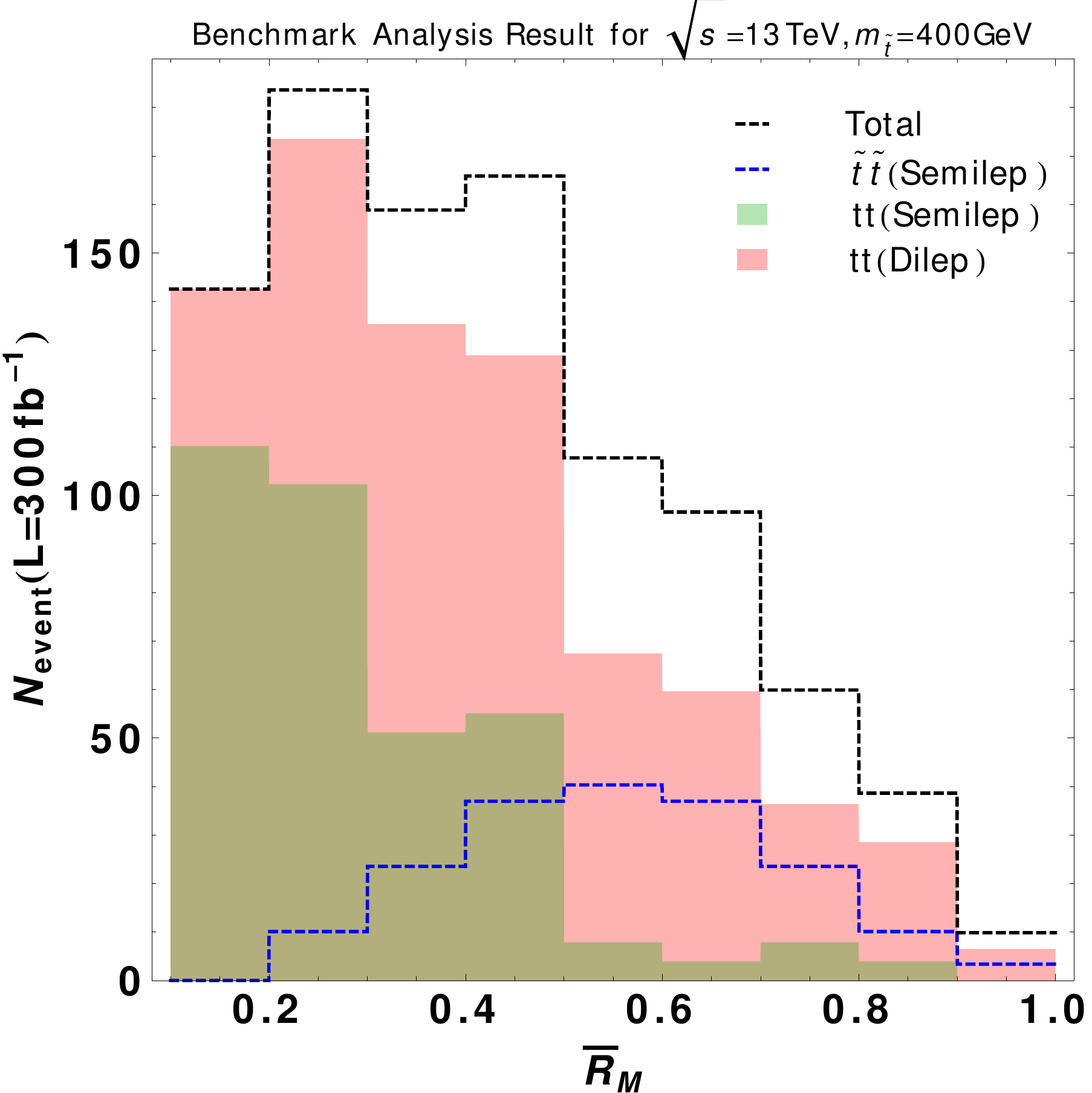}
\caption{Semileptonic decays of the stops.}
\label{benchmark}
\end{subfigure}
\begin{subfigure}[b]{0.4\textwidth}
\includegraphics[width=\textwidth]{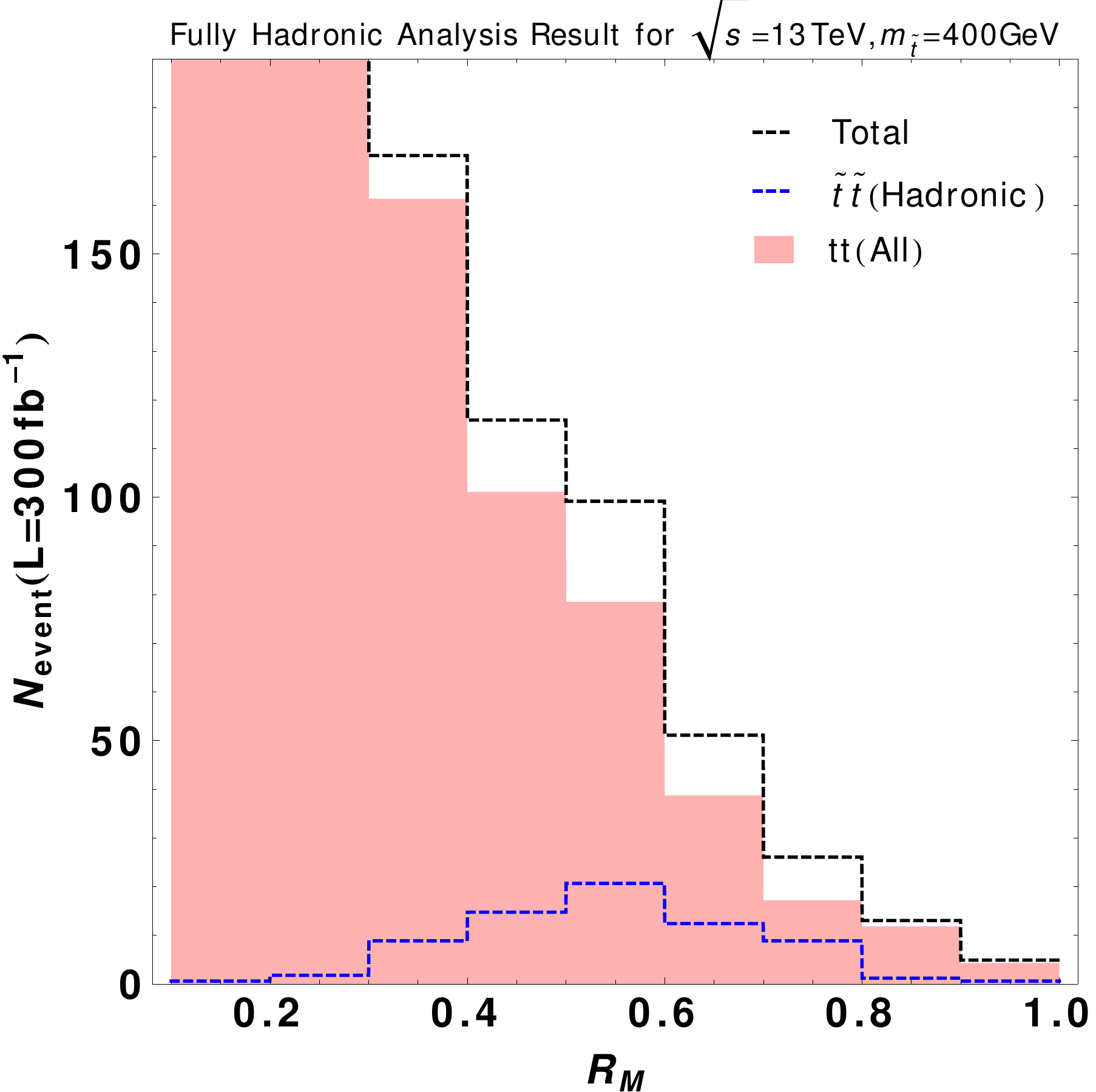}
\caption{Hadronic decays of the stops.}
\label{hadronic}
\end{subfigure}
\caption{\textbf{Left:} the $\bar{R}_M$ distribution for the signal and the backgrounds at $m_{\tilde{t}}=400$ GeV and $m_{\chi}=226.5$ GeV for semileptonic stop decays. \textbf{Right:} the $R_M$ distribution for the same benchmark for hadronic stop decays. The red regions include contributions from the hadronic and semileptonic decays of the $t\bar{t}$ background.}
\end{figure}

Fig.~\ref{benchmark} displays the $\bar{R}_M$ distributions of the signal and backgrounds that pass through all the cuts. 
As shown by the figure, the number of $t\bar{t}$ semileptonic events falls rapidly as $\bar{R}_M$ increases, whereas the dileptonic background shows less a falling trend. This is as expected, because in the case of dileptonic $t\bar{t}$ event, the presence of two neutrinos increases the amount of missing momentum and therefore is more likely to mimic the MET from a genuine semileptonic $\tilde{t}\bar{\tilde{t}}$ event.
Moreover, the signal alone shows a prominent peak at around 0.6. This approximately agrees with the expected $\bar{R}_{M}$ value for the stop decays, which is given by $R_{M}^{\text{theory}}\equiv m_\chi /m_{\tilde{t}} =(400-173.5)/400 \approx0.57$ in this case. However, the total signal plus background is contaminated by the $t\bar{t}$  dileptonic background, therefore has less significant features. After applying a cut at $1 \geq \bar{R}_M \geq R_{M}^{\text{theory}}-0.15 \approx 0.42$, we get 144 signal yields and 250 background yields for 300 fb$^{-1}$ integrated luminosity. A simple estimate of the signal significance using~\cite{Stats1}
\begin{equation}
\sigma =\sqrt{2\left[(S+B)\log\left(1+S/B\right)-S\right]}
\label{eq:significance}
\end{equation}
gives 8.4$\sigma$.

\begin{figure}[t]
\captionsetup{justification=raggedright,
singlelinecheck=false
}
\centering
\includegraphics[scale=0.4]{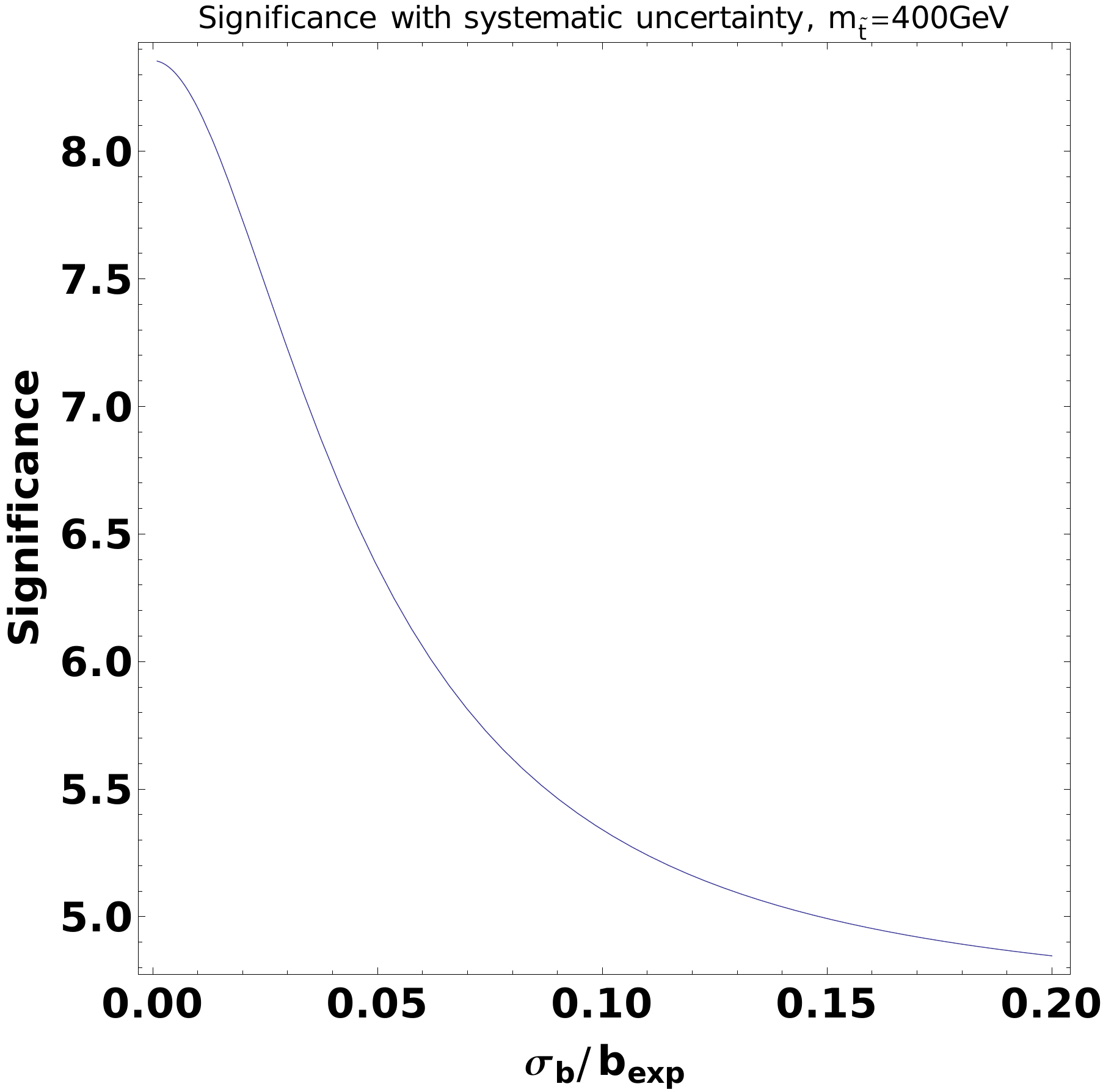}
\caption{Significance as a function of the fractional background uncertainty for the case study.}
\label{uncertain}
\end{figure}

A more sophisticated estimate to take into account the difference between the shapes of the signal and background can be obtained by the likelihood method.
The likelihood ratio between signal plus background hypothesis and background only hypothesis is given by
\begin{equation}
Q\equiv \frac{\mathcal{L}(\{x\};\{s+b\})}{\mathcal{L}(\{x\};\{b\})},\quad \quad \mathrm{where}\,\,\, \mathcal{L}(\{x\};\{\mu\})\equiv\prod_i\frac{\mu_i^{x_i}e^{-\mu_i}}{x_i!}.
\end{equation}
The theoretical predictions for each bin $\{s\}$ and $\{b\}$ are taken from the MC simulation, i.e. Fig.~\ref{benchmark}. The observed number of events for each bin $\{x\}$ are taken to be the simulated signal plus background events rounded to the nearest integers.\footnote{In principle one should do a pseudo experiment to get the ``observed'' number of events.} The significance is given by $\sqrt{2\mathrm{Log}(Q)}$. For our case it gives 8.45, similar to the result of the simple cut analysis using Eq.~(\ref{eq:significance}), which corresponds to treating the entire region after the cut as one bin. The likelihood method also allows us to include uncertainties in the background normalization. Assuming that the background in each bin is a normal distribution around its central value $b$ with an uncertainty $\sigma_b$, the new expression for the likelihood ratio is obtained by
\begin{equation}
Q'= \frac{\int\mathcal{L}(\{x\};\{s+b'\})\mathcal{P}(b')\mathrm{d}b'}{\int\mathcal{L}(\{x\};\{b'\})\mathcal{P}(b')\mathrm{d}b'},\quad\quad \mathrm{where}\,\,\, \mathcal{P}(b')= \frac{1}{\sqrt{2\pi}\sigma_b}e^{-(b-b')^2/2\sigma_b^2}.
\end{equation}
The integration can be done numerically and the upper and lower bounds of the integration are chosen to be $b\pm5\sigma_b$.
The significance obtained is a function of the fractional uncertainty $\sigma_b/b_{exp}$, as shown in Fig.~\ref{uncertain}, where we see that the significance can still maintain as high as 5$\sigma$ even with a 20\% uncertainty in the background normalization.

To compare our result with the study based on fully hadronic final states, we repeat the analysis done by Ref.~\cite{Liantao} for the benchmark. Fig.~\ref{hadronic} shows the result obtained after applying the selections adopted in Ref.~\cite{Liantao}:
$p_T (J_{\text{ISR}}) > 700$~GeV,  
3 sub-leading jets with $p_T>60$~GeV, one or more $b$-tags, 
$|\Delta \phi (J_{\text{ISR}}-\text{MET})-\pi|<0.15$, and 
$|\Delta \phi (\text{jet}-\text{MET})|>0.2$ where ``jet'' is any of the 4 leading jets.
 As can be seen from Figs.~\ref{benchmark} and \ref{hadronic}, the semileptonic stop decays benefit from requiring a lepton in the final state, therefore enjoy a smaller background compared to the hadronic stop decays. After applying a cut at $1 \geq R_M \geq 0.42$, we get 57 signal yields and 232 background yields for the fully hadronic channel, which roughly corresponds to a 4$\sigma$ significance.

\section{Results at LHC 13 TeV}
\label{sec:results}

\begin{figure}[th]
\captionsetup{justification=raggedright,
singlelinecheck=false
}
\centering
\includegraphics[scale=0.4]{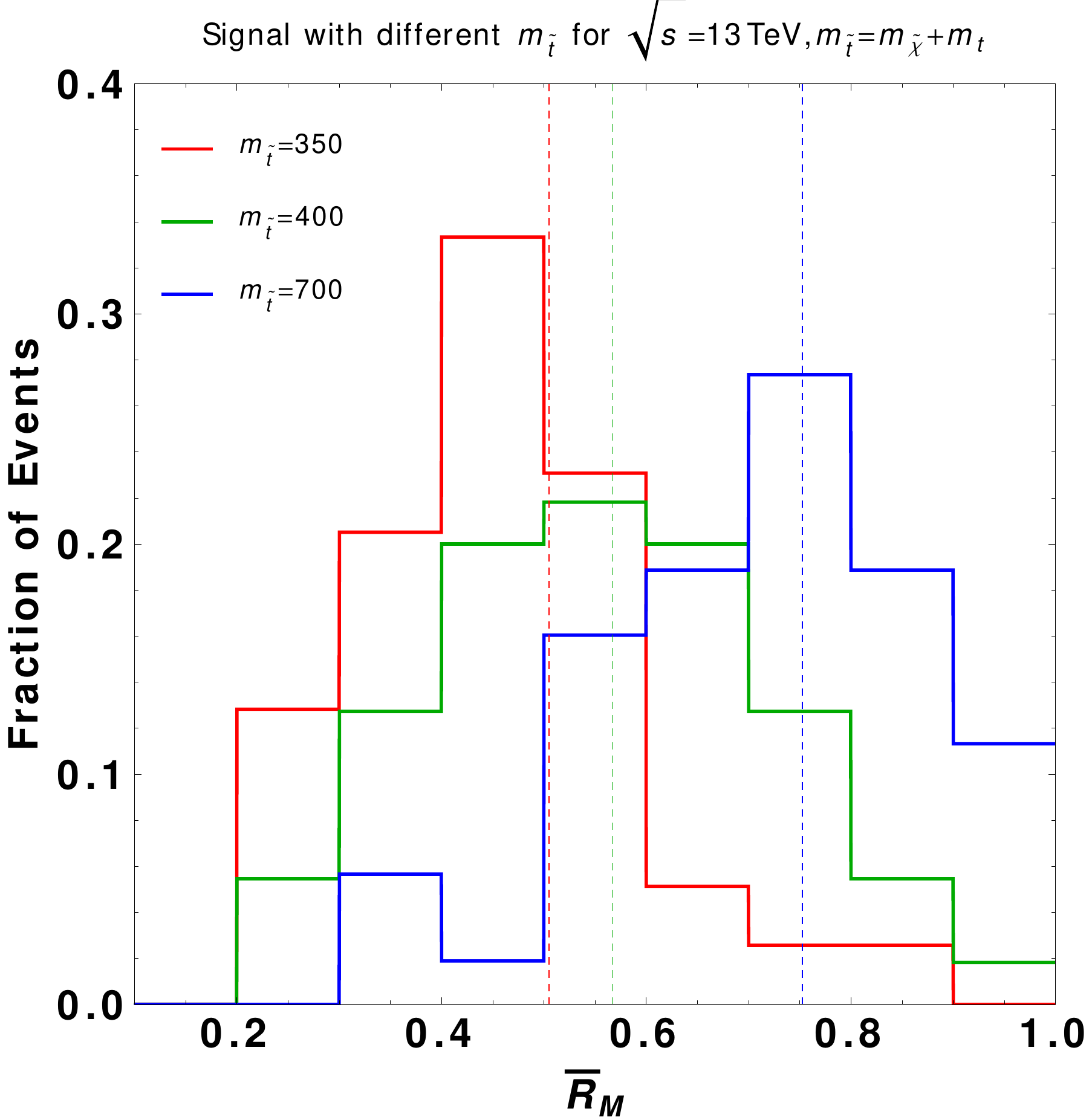}
\caption{The normalized $\bar{R}_M$ distributions of stop signal with $m_{\tilde{t}} = m_t + m_{\tilde{\chi}}$. The Red, Green and Blue curves represent the signals with $m_{\tilde{t}}=350,\, 400,\,  700$ GeV, respectively. The dashed vertical lines are the $R_{M}^{\text{theory}}$ values in these cases.}
\label{movingR}
\end{figure}

We have demonstrated that our method can produce a large signal significance for a 400 GeV stop in the compressed region with 300 $\textrm{fb}^{-1}$ integrated luminosity in the case study. To check how well the $\bar{R}_M$ distribution tracks $R_{M}^{\text{theory}}$ as the masses of the stop and neutralino vary, we perform a series of analyses similar to the benchmark study along the $m_{\tilde{t}} = m_t + m_{\tilde{\chi}}$ line. As examples, the normalized $\bar{R}_M$ distributions for signals at $m_{\tilde{t}} = 350, 400, 700$ GeV are plotted in Fig.~\ref{movingR}.
It is clear that the peaks of the $\bar{R}_M$ distributions follow $R_{M}^{\text{theory}}$ closely along this compressed line.

It is also interesting to see whether or how well this method can work when the stop and neutralino mass difference deviates from the top mass, therefore violating the kinematic assumptions governing Eqs.~(\ref{eq2})-(\ref{eq5}).
When the mass gap between $\tilde{t}$ and $\tilde{\chi}$ is larger than $m_t$, the top quarks and their decay products will still be on shell, therefore Eqs.~(\ref{eq3})--(\ref{eq5}) hold. However, the neutralinos would no longer be static in the rest frame of the stops. As a result, the sum of their momenta may no longer be strictly antiparallel to $J_{\text{ISR}}$, thus our assumption that the neutrino is solely responsible for $\slashed{p}^{\perp}_{T}$ (Eq.~(\ref{eq2})) is less valid. The $\bar{R}_M$ value obtained by solving these equations will be smeared by the error in Eq.~(\ref{eq2}) and the smearing is estimated to be
\begin{equation}
\Delta \bar{R}_M \lesssim \frac{\sqrt{2m_t (m_{\tilde{t}}-m_{\tilde{\chi}}-m_t)}}{p_{T(J_{\text{ISR}})}}\ .
\end{equation}
On the other hand, when the stop is lighter than the sum of $m_t$ and $m_{\tilde{\chi}}$, it will decay via the virtual top quark. Since the LSP $\tilde{\chi}$ is a stable particle, it must be produced on shell. The virtual top will be almost static in the rest frame of the stop, therefore Eq.~(\ref{eq2}) still holds. Eqs.~(\ref{eq3}), (\ref{eq4}) also hold, too, for $W$ and $b$ being on shell. In theory the right hand side of Eq.~(\ref{eq5}) should be modified to $(m_{\tilde{t}}-m_{\tilde{\chi}})^2$ instead of $m^2_t$. In the vicinity of $m_{\tilde{t}}=m_{\tilde{\chi}} + m_t$, Eq.~(\ref{eq5}) approximately holds and $\bar{R}_M$ solved by Eqs.~(\ref{eq2})--(\ref{eq5}) could still be effective.

\begin{figure}[t]
\captionsetup{justification=raggedright,
singlelinecheck=false
}
\centering
\includegraphics[scale=0.4]{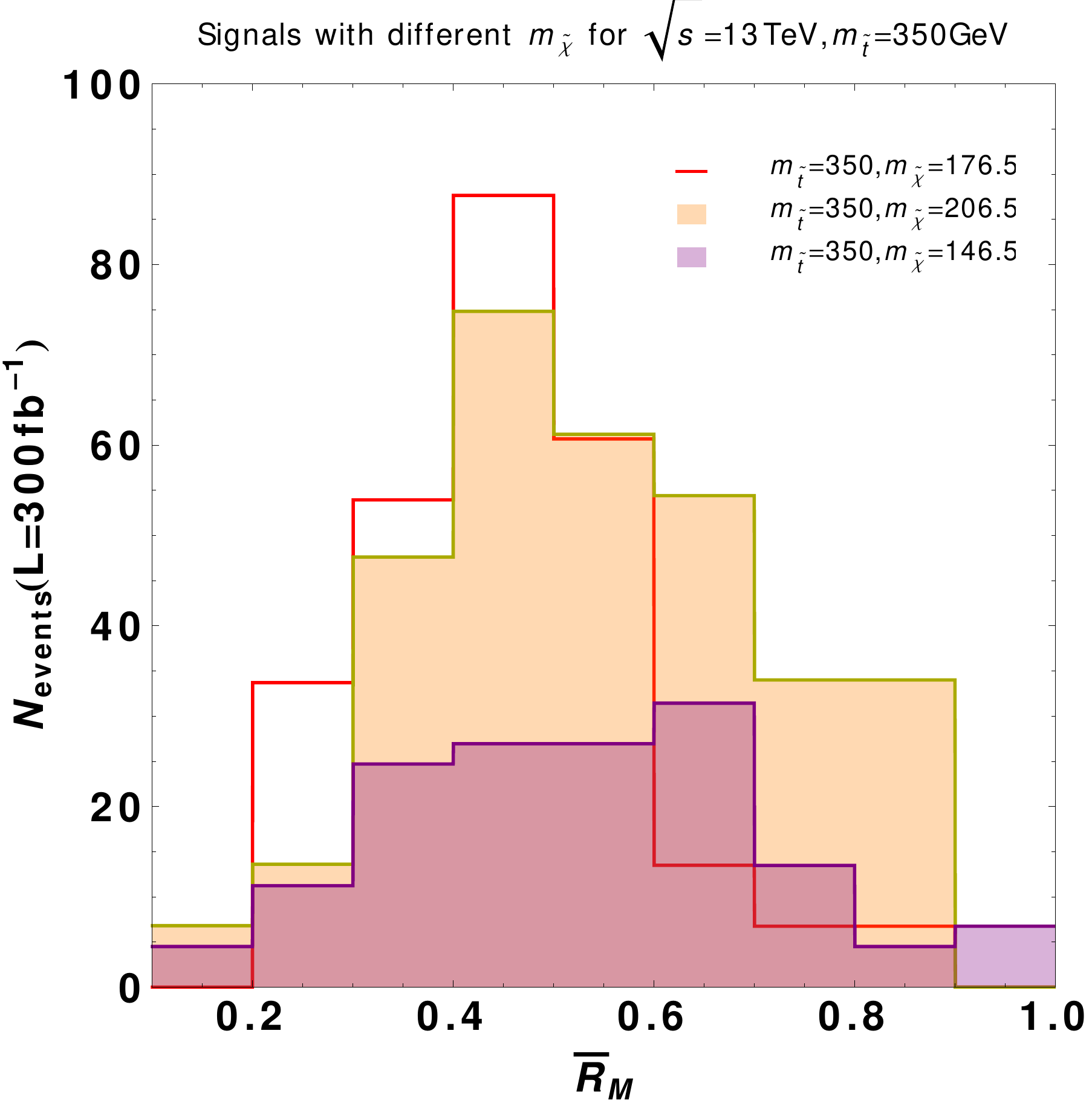}
\caption{Number of signal events distributed along $\bar{R}_M$ with $m_{\tilde{t}}=350$ GeV but different $m_{\tilde{\chi}}$s.}
\label{different_chi}
\end{figure}

To demonstrate how the deviations affect the retrieved $\bar{R}_M$, we compare the number of signal events obtained after employing the same kinematic cuts at $m_{\tilde{t}}=350$ GeV but different $m_{\tilde{\chi}}$s in Fig.~\ref{different_chi}. 
For the case $ m_{\tilde{\chi}}=206.5\, \mathrm{GeV}$ (light orange), the peak stays at the same place as the case of $ m_{\tilde{\chi}}=176.5\, \mathrm{GeV} (= m_{\tilde{t}}- m_t)$ but the distribution is distorted towards larger $\bar{R}_M$ as $m_{\tilde{\chi}}/m_{\tilde{t}}=0.59$ is larger in this case. Even though the peak does not occur at the $R_M^{\text{theory}}$, the signal significance can still be high, since the background distribution diminishes at large $\bar{R}_M$s. In contrast, the $m_{\tilde{\chi}}= 146.5$ GeV case (pink) loses much more events compared to the others and the peak is smeared. This implies that the kinematic assumption of Eq.~(\ref{eq2}) is less appropriate in the scenario $m_{\tilde{t}} - m_{\tilde{\chi}} > m_t$. These results suggest that our method is more powerful in the region $m_{\tilde{t}} - m_{\tilde{\chi}} \lesssim m_t$.

\begin{figure}[t]
\captionsetup{justification=raggedright,
singlelinecheck=false
}
\centering
\includegraphics[scale=0.7]{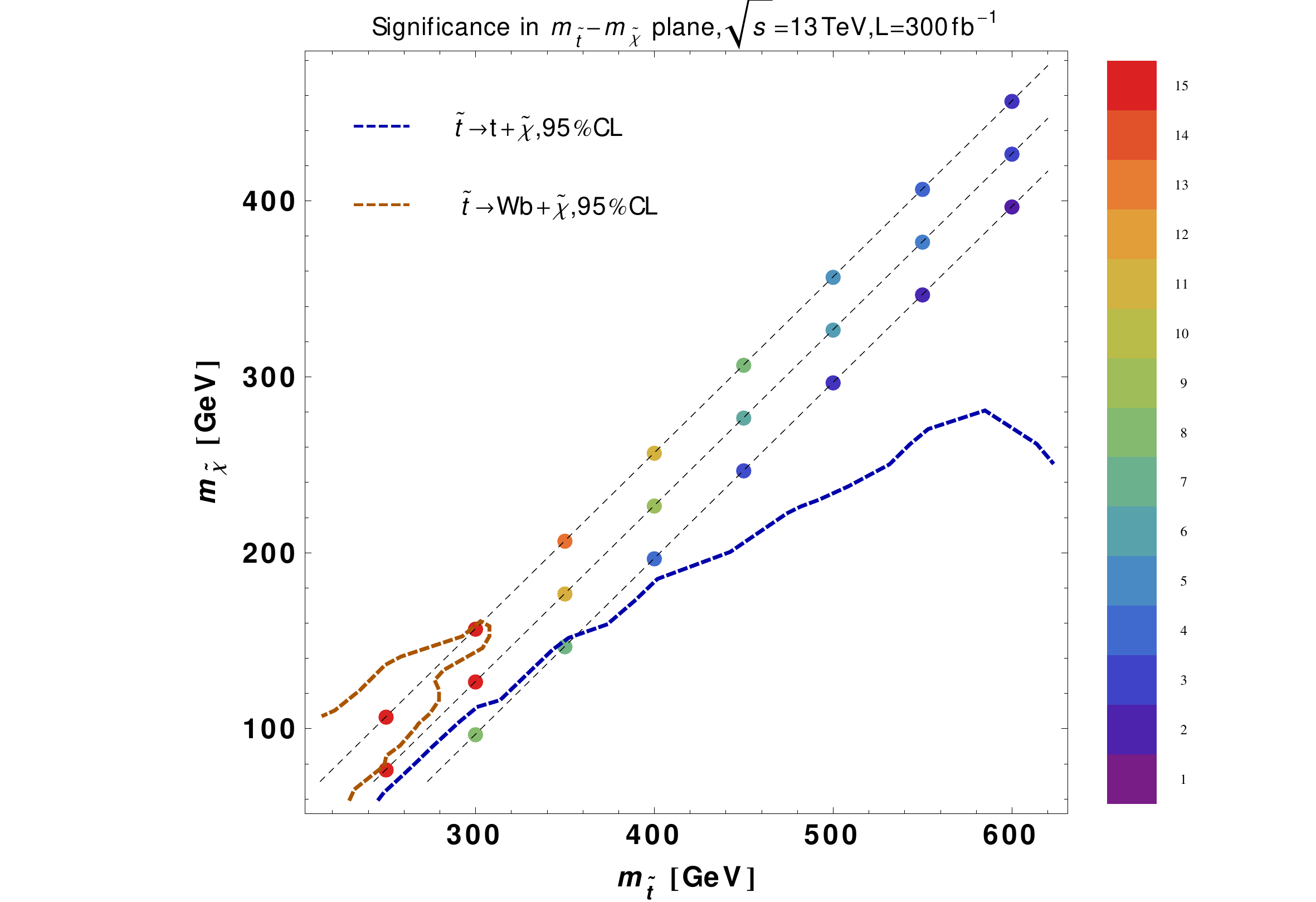}
\caption{The significance gained by semileptonic analysis around the center of compressed region. Three diagonal dashed lines indicate when $m_{\tilde{t}}- (m_{\tilde{\chi}} + m_t) = 30, 0, -30$ GeV. It can be seen clearly that our method can cover a wide mass range when $m_{\tilde{t}}-m_{\tilde{\chi}} \lesssim m_t$. The curves are the exclusion limits from ATLAS~\cite{Aad:2015pfx}}
\label{mass_scan}
\end{figure}
Finally a scan of ($m_{\tilde{t}},\, m_{\tilde{\chi}}$) in the compressed region is performed based on our method. The result is shown in Fig.~\ref{mass_scan}. The scan is done along the $m_{\tilde{t}}- (m_{\tilde{\chi}} + m_t) = 30, 0, -30$ GeV lines. The significances are calculated using the simple expression Eq.~(\ref{eq:significance}) after applying the selections discussed in Sec.~\ref{sec:case}, which are: 
\begin{compactitem}
\item $p_{T(J_{\text{ISR}})} > 475$ GeV.
\item The second and third hardest jets with $p_T > 60$ GeV.
\item MET $>200$ GeV.
\item $|\phi_{J_{\text{ISR}}}-\phi_{\text{MET}}| \geq 2$.
\item $|\phi_{\text{lepton}}-\phi_{\text{MET}}| \geq 0.9$.
\item $R_M^{\text{theory}}-0.15 \leq \bar{R}_M \leq 1, \quad R_M^{\text{theory}} =  \frac{m_{\tilde{\chi}}}{m_{\tilde{t}}}$ for $m_{\tilde{t}} \leq m_{\tilde{\chi}}+ m_t$. For $m_{\tilde{t}} > m_{\tilde{\chi}}+ m_t$ cases, $ R_M^{\text{theory}} =  \frac{m_{\tilde{t}}-m_t}{m_{\tilde{t}}}$ in order to prevent it from being too small.
\end{compactitem}
In Table \ref{sig}, we present the significances for all the points we studied in the compressed region. As expected, the $m_{\tilde{t}}-(m_{\tilde{\chi}} + m_t) = -30$ GeV line achieves as great significances as the $m_{\tilde{t}}-(m_{\tilde{\chi}} + m_t) = 0$ line. They even perform better for lighter stops. This is because the $R_M^{\text{theory}}$ is higher for heavier $m_{\tilde{\chi}}$ given the same $m_{\tilde{t}}$, which means more events distributed at larger $\bar{R}_M$ values, where the SM backgrounds are smaller. On the other hand, the $m_{\tilde{t}}-(m_{\tilde{\chi}} + m_t) = 30$ GeV line performs far worse compared to the other two lines. Overall, the final significances of the three lines agree well with our earlier observation from Fig.~\ref{different_chi}.

\begin{table}[h]
\captionsetup{justification=raggedright,
singlelinecheck=false
}
\begin{center}
\caption{Significances obtained from the $\bar{R}_M$ analysis for the compressed region stops, assuming an integrated luminosity of 300 $\textrm{fb}^{-1}$ at LHC 13 TeV.}
\label{sig}
\begin{tabular}{|l|c|c|c|c|c|c|c|c|c|}
\hline 
$m_{\tilde{t}}$ (GeV)& 250 & 300 & 350 & 400 & 450  & 500 & 550 & 600   \\ 
\hline 
$\sigma_{m_{\tilde{t}}-(m_{\tilde{\chi}} + m_t) = 0}$&19.7 & 15.8 & 11.0 & 8.4 & 5.8  & 5.1 & 3.8 & 2.1   \\ 
\hline 
$\sigma_{m_{\tilde{t}}-(m_{\tilde{\chi}} + m_t) = -30}$& 22&19&13&11&7.2&4.7& 3.1& 1.7 \\ \hline
$\sigma_{m_{\tilde{t}}-(m_{\tilde{\chi}} + m_t) = 30}$& -- &7.6&5.3&3.3&2.4&1.7&1.3 & 0.9 \\ \hline
\end{tabular} 
\end{center}
\end{table}

\section{Conclusions}
\label{sec:conclusions}

In this paper we investigate the stop search from the direct stop pair production in the compressed region, using the semileptonic decay mode. With a hard ISR jet, the neutralinos from the stop decays are boosted in the opposite direction to the ISR jet, producing MET antiparallel to the ISR. Although the neutrino from the leptonic $W$ decay generates additional MET, its momentum can be reconstructed by assuming that the MET transverse to the direction of the ISR jet is entirely coming from the neutrino, together with the mass-shell conditions. The MET due to the neutralinos can be obtained after subtracting the neutrino contribution, and its ratio to the ISR momentum provides a useful kinematic variable $\bar{R}_M$ for the stop search in the compressed region. With proper kinematic cuts, $\bar{R}_M$ distribution for the stop signal events shows a prominent peak around the theoretical value of $m_{\tilde{\chi}}/m_{\tilde{t}}$. The dominant backgrounds are semileptonic and dileptonic top quark pair events. They have a falling distribution in $\bar{R}_M$ and hence may be distinguished from the signal. Other backgrounds are highly suppressed by our event selections and the real solution requirement of the kinematic equations.

Compared with the fully hadronic decay channel, our method for the semileptonic channel requires more sophisticated kinematic reconstruction, but suffers from less SM backgrounds. As a result, we show that the semileptonic channel can have a better reach than the fully hadronic channel along the compressed line $m_{\tilde{t}}-m_{\tilde{\chi}} = m_t$. For 300 $\text{fb}^{-1}$ integrated luminosity at LHC 13 TeV, the semileptonic channel can have a discovery reach of the stop mass up to about 500 GeV, in comparison to $\sim 400$ GeV for the fully hadronic channel. Even though our kinematic equations are strictly valid only for  $m_{\tilde{t}}-m_{\tilde{\chi}} = m_t$, as long as the deviations from this relation is small, the kinematic reconstruction still works pretty well. The reach is somewhat degraded for $m_{\tilde{t}}-m_{\tilde{\chi}} > m_t$ but not for $m_{\tilde{t}}-m_{\tilde{\chi}} \lesssim m_t$.

The stops hold the key to the SUSY solution to the hierarchy problem. Their searches are indisputably important. The traditional stop searches are ineffective for a spectrum of $m_{\tilde{t}}-m_{\tilde{\chi}} \approx m_t$. By resorting to a hard ISR jet, one can construct kinematic variables which can be used to distinguish the stop signal from the very similar SM top backgrounds. Future LHC runs will have a significant coverage of the stop mass even in the compressed region, probing the heart of natural SUSY.

\section*{Acknowledgments}
We would like to thank Zhangqier Wang for discussion on the likelihood method. This work is supported in part by the US Department of Energy grant DE-SC-000999. 


\end{document}